\documentstyle[12pt,lathuile]{article}
\begin{document}
\title{ 
DIRECT AND INDIRECT DETECTION OF WIMPS
}
\author{
  O.~Martineau for the EDELWEISS collaboration \\
{\em Institut de Physique Nucl\'eaire de  Lyon,}\\
{\em IN2P3-CNRS, Universit\'e Claude Bernard Lyon I}\\
{\em 4, rue Enrico Fermi, 69622 Villeurbanne Cedex, France}  \\
}
\maketitle
\baselineskip=14.5pt
\begin{abstract}
We present here the principles of detection of Weakly Interacting Massive Particles, which could represent 
a large contribution to Dark Matter. A status of the experimental situation 
is given both for indirect and direct detection. In particular, the DAMA claim for a WIMP signal is
confronted to the recent results of the CDMS and EDELWEISS experiments. We conclude by comparing direct and 
indirect search sensitivities.
\end{abstract}
\baselineskip=17pt
\newpage
\section{Introduction}
Recent astrophysical observations of the Cosmic Microwave Background\cite{boom}\cite{max} and high-redshift
Supernovae\cite{SN1a1}\cite{SN1a2} tend to favor a Universe where matter is mainly dark\cite{ana} (that is to
say, not emitting electromagnetic radiation). Dark Matter may be present at galactic scale under the
form of a halo composed of weakly interacting, massive -more than 10 GeV/c$^2$- particles, the WIMPs.
These hypothetical relic particles from an early phase of the Universe could be the Ligthest
Supersymmetric Particles of SUSY models (probably the neutralino)  under the hypothesis of R-parity 
conservation.
Two techniques have been proposed to put WIMPs in evidence: indirect and direct
detection. Indirect detection consists in looking for by-products of WIMP-pair 
annihilation in cosmic-rays, while direct search looks for an interaction of a WIMP from the galactic halo with a 
terrestrial target. We will present and compare principles and results of both techniques here. A more
complete review of the subject is given elsewhere\cite{houches}\cite{Klap}.
\section{Indirect detection}
In Minimum Supersymmetric Models, neutralinos are Majorana particles and therefore pair annihilations 
are possible and could be detected through their by-products: $\gamma$ rays, positrons, antiprotons or 
neutrinos, to list the ones of experimental interest.
Here, we will focus on neutrinos, since it seems to be, at this point, the candidate for which the 
experimental signature would be the most telling.
\subsection{Principle of neutrino detection}
Muon neutrinos can be produced by WIMP-pair annihilation directly or through decay of a lepton or a hadron pair
with a typical energy of the order of the third to the half of the mass of the WIMP, and thus in the tens to
hundreds GeV/c$^2$ range.
Their most significant production sites would be the center of the Earth or of the Sun.
Indeed, they are the only regions where the WIMP density 
-enhanced by gravitationnal capture- might be high enough to give consequent reaction rates, though Gondolo
and Silk\cite{gond} have shown that there may also be a significant amplification of the WIMP-annihilation 
rate at the center of our Galaxy.\\
A neutrino can be detected through the upward-going muon produced in the core
of our planet in a charged-current reaction. Having the information on the
neutrino's source (the muon is pointing to it) and a lower limit on its energy (given by the energy of the
muon), it is possible to select the high energy
neutrinos from WIMP-pair annihilation originating from the Sun or the center of the Earth. A large part of 
the background noise is thus discarded (solar neutrino energies are for
example typically of the order of 20 MeV/c$^2$), and the only remaining sources of background are atmospheric 
neutrinos produced on the other side of our planet and neutrinos produced by cosmic-ray interaction in the 
corona of the Sun.
\subsection{Results and perspectives}
The first results on indirect WIMP search came in the eighties from detectors developed for other 
purposes, like the study
of the proton decay (IMB), or the study of solar or atmospheric neutrinos (Baksan, Kamiokande, MACRO, etc).
They were thus not totally adapted to WIMP detection (detection areas of the order of $10^3$ m$^2$), 
but already provided limits on the upward-going muon flux\cite{baksan}\cite{MACRO} interesting enough to 
reject the MSSM models for WIMPs yelding the largest neutralino rates.\\
High-energy neutrino telescopes (ANTARES\cite{ANTARES} or AMANDA\cite{AMANDA} for example) are now under 
development. Large sizes should be achievable (up to the km$^2$, which corresponds to a factor $10^4$ 
improvement with regard to previous neutrinos detectors), since a  natural medium is used as their detecting 
volume (water of the Mediterranean sea for ANTARES, ice of the Antarctic for AMANDA). Despite a higher
energy threshold, their sensitivity to WIMP annihilation in the Sun or in the Earth  is expected to be 
improved by the same $10^4$ factor, which corresponds
 to a muon flux of the order of 10 per km$^2$ per year for an exposure of 10 km$^2$.yr\cite{berg}. 
This sensitivity should allow these experiments to test a large part of the domain allowed by MSSM models 
in the coming years.
\section{Direct detection}
\subsection{Principle}
Another possibility to put WIMPs in evidence consists in looking for the scattering of a WIMP from the galactic 
halo on a target detector placed on Earth, in which it would produce a nuclear recoil. This type of search 
can be readily extended to WIMP models beyond the MSSM since the only necessary
condition for detection is a non-zero WIMP-nucleon cross-section. In the hypothesis 
where the WIMP is the MSSM  neutralino, Goodman and Witten have shown\cite{Good}
that it could couple to a quark of the scattered nucleus via two mechanisms: spin-dependent (Z-boson or squark 
exchange) or spin-independent (Higgs bosons or squark exchange). It can be shown\cite{JKG} that the 
spin-dependent cross-section $\sigma_{SD}$ is proportional to the nucleus spin J of the
target, while the spin-independent cross-section $\sigma_{SI}$ is grossly proportional to the square of the atomic 
number A of the nucleus. It follows that $\sigma_{SI}>\sigma_{SD}$ for $A>30$ 
(which corresponds to a large majority of the targets) and that the interaction rate per kg of matter varies between 
1 event per day to one 
per decade, depending on model parameters and target nuclei. 
The two main requirements for direct search detectors are thus low radioactive background rates and, since 
WIMPs induced nuclear recoils are below 100 keV, low energy thresholds.\\
A positive signature could come from the annual modulation of the event rate in the detector. Indeed the 
relative velocity of the Earth with regard to the galactic halo varies annually with the rotation of the 
Earth around the Sun. Thus there should be a small variation of approximately 5\% in the WIMP event rate in 
the detector. A significant experimental signature would nevertheless require large target masses 
($\geq$ 100 kg) and an excellent stability of the detector performances -better than the percent-, even under
the extremely favorable hypothesis of the absence of background and for the SUSY models yelding the largest
neutralino rates\cite{houches}.\\
Several different solutions have been proposed to fulfil the heavy constraints of direct detection. We
will present here those giving at this point the best results, even if some other innovating techniques 
seem promising\cite{houches}.
\subsection{Classical detectors}
Historically the first type of detectors used for direct detection were germanium ionisation detectors at 
liquid nitrogen temperature. The interaction is detected through the collection of the charge of the 
electron-hole pairs created in the crystal. 
Thanks to years of development in the fields of $\gamma$ and $\beta$ spectroscopy and high performance germanium
purification technique, Ge diodes can reach excellent energy resolutions (typically 1 keV full width
half maximum for 300 keV deposited\cite{HDMS}) and the lowest total event rate of all direct search experiments
(0.042 event/kg/keVee/day between 15 and 40 keV recoil\cite{HeidMos}).
Nevertheless, it is not possible to discriminate nuclear recoils (induced by neutrons or WIMPs) from 
electron recoils (induced by $\beta$ and $\gamma$ radioactivity), which is the dominant 
background. Therefore, the experiments using this technique (HDMS\cite{HDMS}, IGEX\cite{IGEX}, etc.) after holding 
the most stringent
limits on WIMP-nucleon cross-section for a long time, seem now to be limited by this absence of rejection, 
even if projects using this type of detectors could remain competitive in the future\cite{genius}.\\
Scintillators are other classical detectors adaptated to WIMP direct detection. Large masses are achievable
(730 kg for Elegant-V\cite{elegant}, 100 kg for DAMA\cite{DAMA98}). A statistical rejection of the 
$\gamma$ background is possible using the different
scintillation time constants between electrons and nuclear recoils (pulse shape discrimination), but this cannot 
be applied at energies just above threshold\cite{DAMA96} (typically below 5 keV), which correspond to the 
most significant part of the data. The best spin-dependent limit on the WIMP-nucleon cross-section has been 
achieved by the DAMA experiment using NaI 
crystals\cite{DAMA96}, thanks to the non-zero spin of the sodium nuclei.
Recently, the DAMA experiment also claimed an annual modulation signal\cite{DAMA} which this group has 
attributed to a WIMP of mass $52\pm^{10}_8$ GeV and a spin-independent WIMP-nucleon cross-section of
$\sigma_n=(7.2\pm^{0.4}_{0.9})\cdot10^{-6}$ pb. 
If combined to their 1996 exclusion data based on pulse shape discrimination\cite{DAMA96}, the most likely 
WIMP mass and WIMP-nucleon cross-section values become respectively $44\pm^{10}_8$ GeV and 
$\sigma_n=(5.4\pm1.0)\cdot10^{-6}$ pb. This result remains controversial\cite{houches}\cite{Gilles} and is 
hardly compatible with the results of other experiments\cite{CDMS}\cite{Ed} using a new type of detectors, 
bolometers.
\subsection{Bolometers}
\subsubsection{Principle and Performances}
\begin{figure}[t]
 \vspace{9.5cm}
\includegraphics{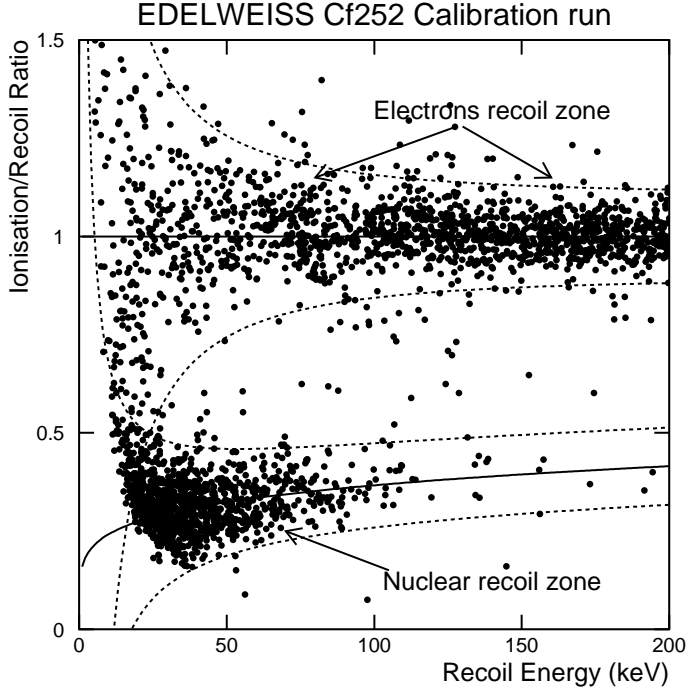}
 \caption{\it
    Plot of the Ionisation/Recoil ratio (Q) against the recoil energy for events recorded in a $^{252}Cf$ 
    calibration run of the Edelweiss 320 g bolometer. The solid lines represent the average Q
    distibution for photons (Q=1 by construction) and for neutrons ($Q=0.16.(E_{Recoil})^{0.18}$), and the
    dashed curves are the limits of the 99.9\% efficency regions for photons and neutrons.
    \label{discri} }
\end{figure}

Bolometers measure the elevation of temperature due to an interaction of a particle in an absorber 
(Saphire\cite{ROSEBUD}, Germanium\cite{rejEd}, or CaWO$_4$\cite{CRESST} for example) by means of a 
thermometric sensor glued to its surface. In principle, 
energy deposits as small as 1 keV result in a measurable elevation of temperature -typically 1 $\mu$K- in a 
100 g detector working at a temperature of 10 mK. The elevation of temperature due to an
energy deposit in the absorber is indeed inversely proportionnal to its heat capacitance, which is very
small at these temperatures. 
Furthermore the fundamental resolution of such bolometers, given by thermodynamic fluctuations in the
energy of the absorber, is in the tens of electron-volt range.\\
In the field of Dark Matter Search, bolometers offer another very attractive feature, since it is possible 
to reject electron recoils with a high efficiency for certain types of absorber. The number of charges 
created in a semiconducting absorber (Germanium or Silicon) by nuclear recoils is indeed 
approximately three times lower than for an electron recoil of the same energy. By
measuring simultaneously the ionisation and the heat signals for every interaction, the CDMS and 
EDELWEISS\cite{rejEd} experiments can thus discriminate $\gamma$ and $\beta$ radioactive backgrounds from 
possible WIMP-induced events with a rejection factor higher than 99\% (see fig.\ref{discri}). 
The ROSEBUD\cite{PdM} and CRESST\cite{CRESST} experiments have shown that the measurement of the 
scintillation light emitted in CaWO$_4$ absorbers also makes this 
discrimination possible. This active rejection of the background explains why bolometers have the lowest 
nuclear recoil rates of all direct search experiments. It is also the reason why they are already competitive 
with the optimized classical detectors mentioned above, although still in their development phases.
\subsubsection{Results}
\begin{figure}[t]
 \vspace{8.5cm}
\includegraphics{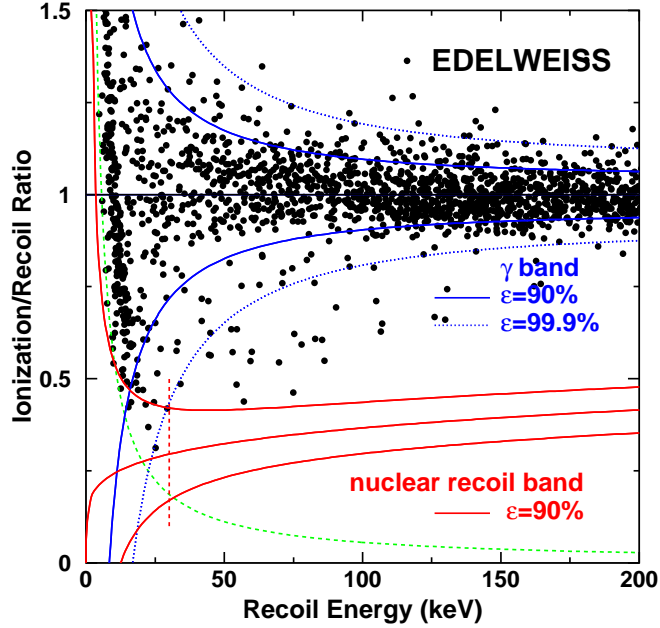}
 \caption{\it
    Plot of the Ionisation/Recoil ratio (Q) as a function of the recoil energy from the data collected in the
    centre fiducial volume of the 320g EDELWEISS detector. Also plotted are the $\pm1.645\sigma$ bands (90\%
    efficiency) for photons and for nuclear recoils. The 99.9\% efficiency region for photons is also shown
    (dotted line). The hyperbolic dashed curve corresponds to 5.7 keV ionisation energy and the vertical dashed
    line to 30 keV recoil energy.
    \label{fond} }
\end{figure}

\begin{figure}[t]
 \vspace{9.0cm}
\includegraphics{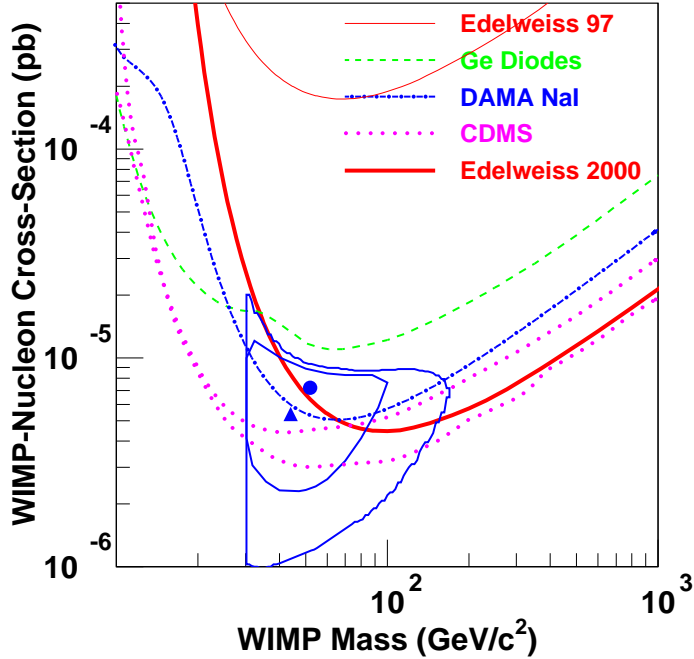}
 \caption{\it
    WIMP-nucleon spin-independent cross-section as a function of Wimp mass. 
    Light solid curve: Limit obtained by EDELWEISS with the 70g bolometer data\cite{rejEd}. 
    Dark solid curve: Limit obtained by EDELWEISS with the
    320g bolometer data\cite{Ed}.
    Dashed curve: combined Ge diode limit\cite{HDMS}\cite{HeidMos}\cite{IGEX}. 
    Dash-dotted curve: 1996 DAMA-NaI0 limit using pulse shape discrimination\cite{DAMA96}. 
    Light dotted curve: CDMS limit without statistical subtraction of the
    neutron background\cite{CDMS}. 
    Dark dotted curve: CDMS limit with statistical subtraction of the
    neutron background\cite{CDMS}. 
    Dark closed contour: allowed region at $3\sigma$ CL for a WIMP r.m.s velocity of
    270 km/s from the DAMA annual modulation data\cite{DAMA}. 
    Filled circle: central value of the previous $3\sigma$ region (WIMP-mass 52 GeV and 
    WIMP-nucleon cross-section $\sigma_n=7.2\cdot10^{-6}$ pb). 
    Light closed contour: allowed region at $3\sigma$ CL from the DAMA annual modulation data\cite{DAMA} 
    when combined to their 1996 exclusion limit\cite{DAMA96} and accounting for the uncertainty on the 
    WIMP velocity (210-330 km/s rms). 
    Triangle: central value of the previous $3\sigma$ region (WIMP-mass 44 GeV and WIMP-nucleon 
    cross-section $\sigma_n=5.4\cdot10^{-6}$ pb) for a WIMP r.m.s velocity of 270 km/s. 
    \label{exclu} }
\end{figure}

Recently, the Cryogenic Dark Matter Search (CDMS) experiment reported results\cite{CDMS} obtained
with three 165 g Ge heat-and-ionisation detectors. In 96 live days of data acquisition in their
shallow site of Stanford (corresponding to 10.6 kg.days), they recorded 13 nuclear recoils in the 10-100keV 
recoil range. This rate is compatible with that expected from a WIMP of mass 52 GeV and a
WIMP-nucleon cross-section of $7.2\cdot10^{-6}$pb. Nevertheless, the presence of 4 multiple-scatter nuclear
recoils in the Germanium detectors and event rates measured with a 100 g Si cryogenic detector tend to favor 
the hypothesis where these 
13 events are due to cosmic-ray induced neutrons. By making this assumption and subtracting the neutron
background, CDMS then obtained a limit 
for the WIMP-nucleon cross section incompatible with the whole 3$\sigma$ region of the DAMA claim at 84\% CL 
(see fig.\ref{exclu}).\\
The French experiment EDELWEISS, based in the underground site of the Laboratoire Souterrain 
de Modane (LSM), is not limited by the cosmic-ray induced neutron background\cite{LSM} so far, and 
may resolve this 
discrepancy. More recently, EDELWEISS has accumulated an effective exposure of 4.53 kg.days\cite{Ed}
(fiducial volume) with a 320 g heat-and-ionisation Ge detector and observed no nuclear recoils in the 
30-200 keV energy range (see fig.\ref{fond}). This excludes at 90\% CL the central value obtained for the WIMP 
signal reported by DAMA with a WIMP-nucleon cross-section $\sigma_n=7.2\cdot10^{-6}$ pb, but not the central 
value of  $\sigma_n=5.4\cdot10^{-6}$ pb, when the 1996 DAMA-NaI0 exclusion limit is taken into account (see 
fig.\ref{exclu}). More data is thus needed by EDELWEISS to test 
the whole DAMA zone. This should come with the installation of three 320 g bolometers in the LSM, scheduled
at the end of this year.\\
\section{Conclusion: indirect vs direct detection}
\begin{figure}[t]
 \vspace{6.0cm}
\includegraphics{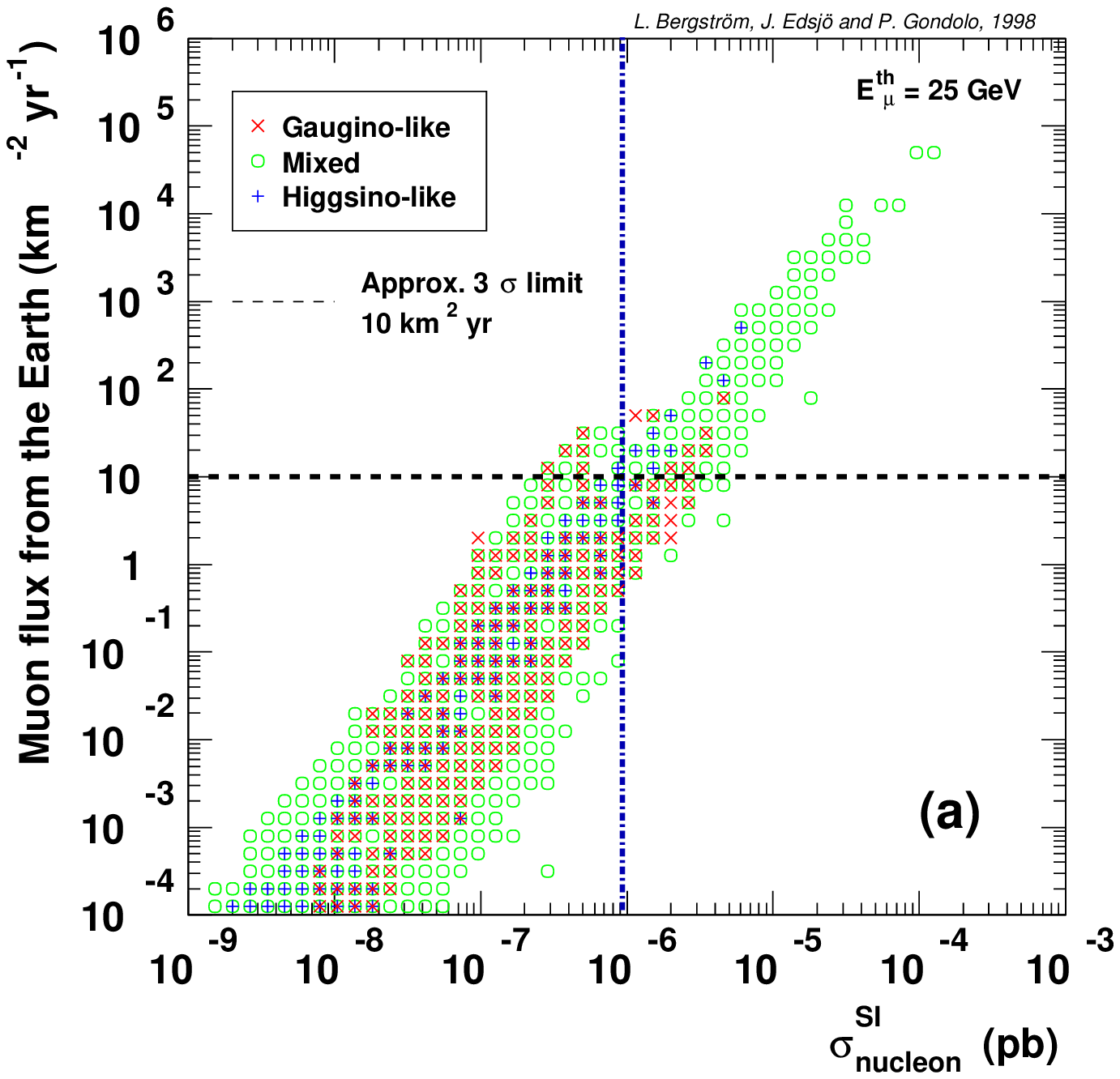}
\includegraphics{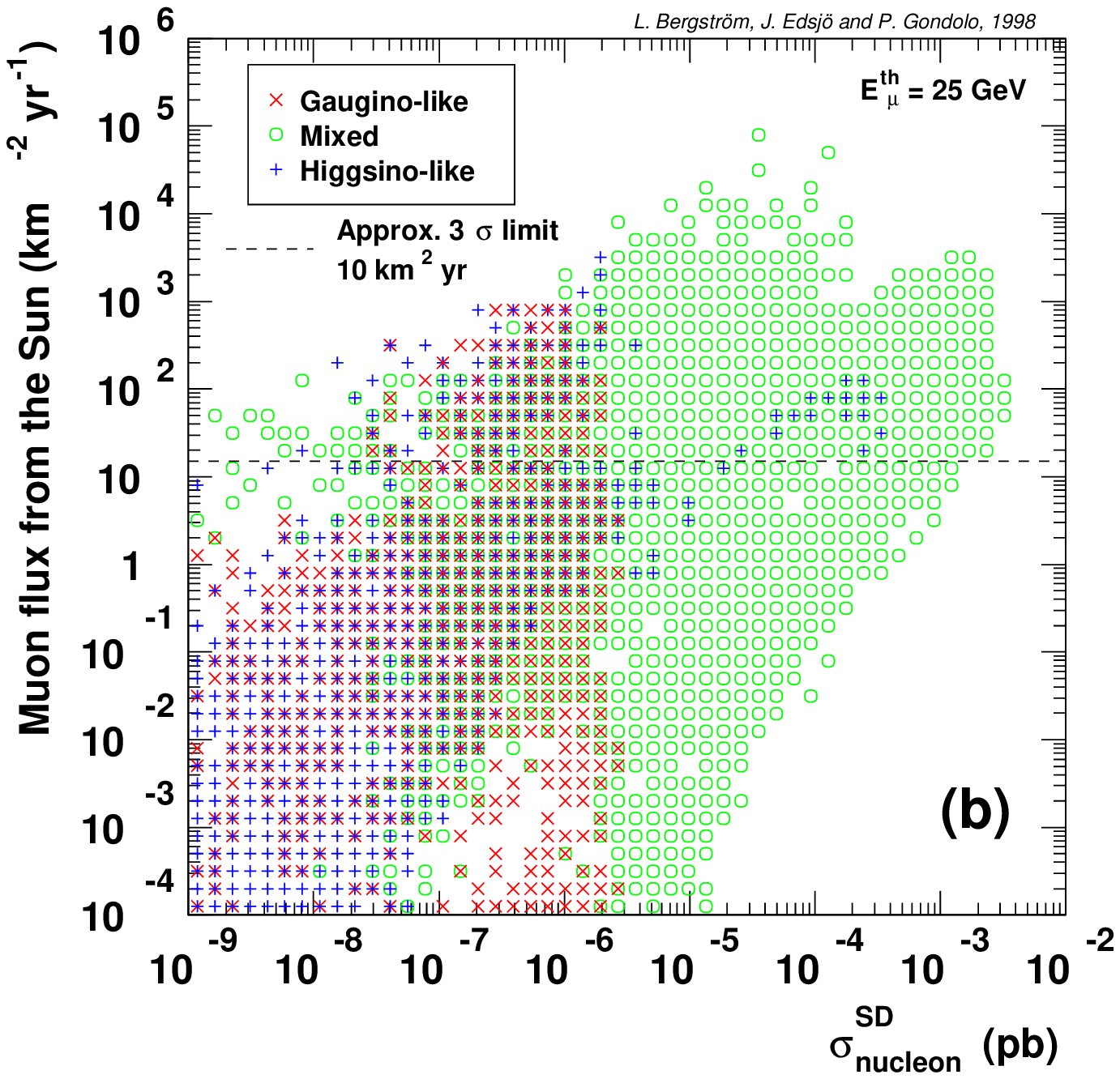}
 \caption{\it
     a): Muon flux from the Earth as a function of spin-independent neutralino-nucleon cross-section expected from
     different models of neutralinos. Dotted line: 3$\sigma$ limit expected on muon fluxes from high-energy 
     neutrinos telescopes with a 10 $km^2$.yr exposure. Dash-dotted line: present best limit on 
     spin-independent neutralino-nucleon cross-section of direct detection experiments\cite{CDMS}.
     b): Muon flux from the Sun as a function of spin-dependent neutralino-nucleon cross-section expected from
     different models of neutralinos. Dotted line: 3$\sigma$ limit expected on muon fluxes from high-energy
     neutrinos telescopes with a 10 $km^2$.yr exposure. The present best limit on spin-dependent
     neutralino-nucleon cross-section of direct detection experiments ($\simeq10^{-1}$ pb\cite{DAMA96}) is too
     large to be shown on this plot. Taken from Bergstr\"om {\it et al.}\cite{berg}.
    \label{comp} }
\end{figure}

It is of course very tempting to compare the sensitivities of direct and indirect WIMP searches. This will be
done here in the MSSM framework used in refs\cite{berg}\cite{berg2} in which no restriction is brought from 
supergravity other than gaugino mass unification.\\
The muon flux due to the annihilation of WIMPs in the Earth depends on the density of the WIMP halo, its
kinematics and the accretion rate of WIMPs in the Sun or the Earth, which itself is determined by the way 
WIMPs scatter off the nuclei composing the core of the Earth (mainly iron or nickel). This scattering is 
mainly spin-independent (since iron and nickel are heavy elements) and it is thus quite similar to the 
interaction taking place in direct detection experiments. Furthermore, the local halo density plays the same role in
both cases. Therefore similar assumptions can be made for indirect and direct detection, and it is
possible to link a given muon flux due to WIMPs annihilation in the core of the Earth to a certain 
WIMP-nucleon cross-section. This was done elsewhere\cite{berg}, and the result, shown in fig.\ref{comp}, is
that the sensitivities for spin-independent interactions expected for the future high-energy neutrinos 
telescopes (10 muons/km$^2$) are roughly
equivalent to the present direct detection experiments limits on WIMPs-nucleon cross section 
($\sigma_{SI}\simeq10^{-6}$ pb).\\
For the case of  muons coming from the Sun, the situation is quite different, since WIMPs interact 
with protons during the accretion phase. In the MSSM framework, it is thus the spin-dependent cross section 
which may become predominant, and the process is quite different from the one taking place in a direct detection 
detector. The comparison is
thus more difficult. Nevertheless, calculations from ref.\cite{berg} show that the future 
high-energy neutrinos telescopes may in this case give better results than the present direct detection experiments 
(fig.\ref{comp}). Indeed the direct detection best spin-dependent
WIMPs-nucleon cross section ($\sigma_{SD}\simeq10^{-1}$ pb\cite{DAMA96}) corresponds to a muon flux from the 
Sun that is several orders of magnitude higher than the 10 muons/km$^2$ expected from high-energy neutrinos 
telescopes, and already much larger than the limits derived from the Kamiokande observations.\\
A similar study led in ref.\cite{JKG} comes to the same conclusion, which tends to prove that both methods are 
complementary and thus worth pursuing.
\clearpage

\begin{thebibliography}{99}
\bibitem{boom}
P. de Bernardis {\it et al}, Nature {\bf404}, 955 (2000).
\bibitem{max}
S.~Hanany {\it et al}, Astrophys. J. {\bf545}, L5 (2000).
\bibitem{SN1a1}
S.~Perlmutter {\it et al}, Astrophys. J. {\bf517}, 565-586 (1999).
\bibitem{SN1a2}
A.G.~Riess {\it et al}, Astron. J. {\bf116}, 1009 (1998).
\bibitem{ana}
A. Balbi {\it et al}, Astrophys. J. {\bf 545} L1-L4 (2000).
\bibitem{houches}
G.~Chardin, Dark Matter: direct detection, in: The primordial Universe (ed. P.~Binetruy, R.~Schaeffer, J.~Silk
and F.~David) (Springer, Berlin, 2000).
\bibitem{Klap}
H.V.~Klapdor-Kleingrothaus and A.~Staudt, Non-accelerator Particle Physics (ed. Institute of Physics
Publishing, Bristol and Philadelphia, 1995).
\bibitem{gond}
P.~Gondolo and J.~Silk, Phys. Rev. Let., {\bf 83}, 1719 (1999).
\bibitem{baksan}
M.M.~Boliev {\it et al}, Nucl. Phys. {\bf B} (Proc. Suppl.), {\bf 48}, 83 (1996).
\bibitem{MACRO}
M.~Ambrosio {\it et al}, Astrophys. J. {\bf 546} 1038-1054 (2001).
\bibitem{ANTARES}
The ANTARES collaboration, astro-ph/9707136 (1997).
\bibitem{AMANDA}
E.~Andres {\it et al}, Astropart.Phys. {\bf 13} 1-20 (2000).
\bibitem{berg}
L. Bergstr\"om {\it et al}, Phys. Rev. {\bf D58} 103519 (1998).
\bibitem{Good}
M.W.~Goodman and E.~Witten, Phys. Rev. {\bf D31} 3059 (1985).
\bibitem{JKG}
G. Jungman, M. Kamionkowski and Kim Griest, Phys. Rev. {\bf 267}, 195 (1996).
\bibitem{HDMS}
L.~Baudis {\it et al}, Phys. Rev. {\bf D63} 022001 (2001).
\bibitem{HeidMos}
L.~Baudis {\it et al}, Phys. Rev. {\bf D59} 022001 (1998).
\bibitem{IGEX}
A.~Morales {\it et al}, Phys. Lett. B {\bf 489} 268 (2000).
\bibitem{genius}
H.V.~Klapdor-Kleingrothaus {\it et al}, to be publ. in proc. World Scientific (2001), and Preprint: hep-ph/0103079.
\bibitem{elegant}
H.Ejiri in: Proc. second International Workshop on the identification of Dark Matter (ed. N.J.C.
Spooner and V. Kudryavtsev), 329 (World Scientist, Singapore, 1999).
\bibitem{DAMA98}
R. Bernabei {\it et al}, Il Nuovo Cim. {\bf A112} 545 (1999).
\bibitem{DAMA96}
R. Bernabei {\it et al}, Phys. Lett. B {\bf 389}, 757 (1996).
\bibitem{DAMA}
R. Bernabei {\it et al}, Phys. Lett. B {\bf 480}, 23 (2000).
\bibitem{Gilles}
G.~Gerbier {\it et al}, astro-ph/9902194 (1999).
\bibitem{CDMS}
R. Abusaidi {\it et al}, Phys. Rev. Lett. {\bf 84}, 5699 (2000).
\bibitem{Ed}
A. Benoit {\it et al}, to be publ. in Phys. Lett. B. and Preprint: astro-ph/0106094.
\bibitem{ROSEBUD}
S.~Cebri\'an {\it et al}, Astropart. Phys. {\bf 15}, 79-85 (2001).
\bibitem{rejEd}
P.~di Stefano {\it et al}, Astropart. Phys. {\bf 14}, 329 (2001).
\bibitem{PdM}
P.~de Marcillac, private communication.
\bibitem{CRESST}
M.~Bravin {\it et al}, Astropart. Phys. {\bf 12}, 107-114 (1999).
\bibitem{LSM}
V.~Chazal {\it et al}, Astropart. Phys. {\bf 9}, 163 (1998).
\bibitem{berg2}
L. Bergstr\"om and P. Gondolo, Astropart. Phys. {\bf 5}, 263 (1996).
\end{thebibliography}
\end{document}